%% file: CIPANP_submit.tex
\documentclass[12pt]{article}
\usepackage{graphicx}

\newcommand{\bctaunutau}{b \to c \tau^- {\bar\nu}_\tau}

\def\BDtaunu{\bar{B} \to D^+ \tau^{-} \bar{\nu_\tau}} 
 
\def\BDlnu{\bar{B} \to D^+ \ell^{-} \bar{\nu_\ell}}
\def\BDstartaunu{\bar{B} \to D^{*+} \tau^{-} \bar{\nu_\tau}}

\def\BDstarlnu{\bar{B} \to D^{*} \ell^{-} \bar{\nu_\ell}}

\def\Rlcr{R_{\Lambda_c}^{Ratio}}

\def\BDtaunu{\bar{B} \to D^+ \tau^{-} {\bar\nu}_\tau}
\def\BDlnu{\bar{B} \to D^+ \ell^{-} {\bar\nu}_\ell}
\def\BDstartaunu{\bar{B} \to D^{*+} \tau^{-} {\bar\nu}_\tau}
\def\BDstarlnu{\bar{B} \to D^{*+} \ell^{-} {\bar\nu}_\ell}

\def\RD{R({D^{(*)}})}

\def\nl{\nonumber\\}

\def\beq{\begin{equation}}
\def\eeq{\end{equation}}
\def\bea{\begin{eqnarray}}
\def\eea{\end{eqnarray}}
\def\nn{\nonumber}
\def\roughly#1{\mathrel{\raise.3ex\hbox
{$#1$\kern-.75em\lower1ex\hbox{$\sim$}}}}

\def\sss{\scriptscriptstyle}

  \def\rr2{{1\over\sqrt{2}}}

\def\.{\!\cdot\!}    \def\:{\cdots}   \def\[{\left[}   \def\]{\right]}
\def\({\left(} \def\){\right)} 

\def\barpk{{\raise.35ex\hbox
{${\sss (}$}}--{\raise.35ex\hbox{${\sss )}$}}}
\def\bbarp{\hbox{$B$\kern-0.9em\raise1.4ex\hbox{\barpk}}}
\def\beq{\begin{equation}}
\def\eeq{\end{equation}}
\def\bea{\begin{eqnarray}}
\def\eea{\end{eqnarray}}
\def\nn{\nonumber}
\def\sss{\scriptscriptstyle}
\def\roughly#1{\mathrel{\raise.3ex\hbox
{$#1$\kern-.75em\lower1ex\hbox{$\sim$}}}}

\def\th1{\theta_{LR}}
\def\th2{\theta_{RL}}

\def \SM{{\rm SM}}

\def \expt{{\rm expt}}

\def \lamc{\Lambda_c}
\def \lbt{\Lambda_b \to \Lambda_c \tau \bar{\nu}_{\tau}}
\def \lbl{\Lambda_b \to \Lambda_c \ell \bar{\nu}_{\ell}}

\def\beq{\begin{equation}}
\def\eeq{\end{equation}}
\def\bea{\begin{eqnarray}}
\def\eea{\end{eqnarray}}
\def\ber{\begin{eqnarray*}}
\def\eer{\end{eqnarray*}}
\def\bwt{\begin{widetext}}
\def\ewt{\end{widetext}}
\def\nn{\nonumber}
\def\roughly#1{\mathrel{\raise.3ex\hbox
{$#1$\kern-.75em\lower1ex\hbox{$\sim$}}}}

\def\order{\lower 1.8ex \hbox{\LARGE\~{}}}

\def\BDtaunu{\bar{B} \to D^+ \tau^{-} {\bar\nu}_\tau}
\def\BDlnu{\bar{B} \to D^+ \ell^{-} {\bar\nu}_\ell}
\def\BDstartaunu{\bar{B} \to D^{*+} \tau^{-} {\bar\nu}_\tau}
\def\BDstarlnu{\bar{B} \to D^{*+} \ell^{-} {\bar\nu}_\ell}

\def\bsmumu{ b \to  s \mu^+ \mu^-}

\def\RD{R({D^{(*)}})}

\def\RK{{R_{K^{}}}}

\def\RKstar{{R_{K^{(*)}}}}

\newcommand\pubnumber{UMISS-HEP-2018-02}
\newcommand\pubdate{\today}

\def\napoli{Department of Physics and Astronomy\\
Department of Physics and Astronomy, University of Mississippi, 108 Lewis Hall, Oxford, MS 38677, USA}
\def\support{\footnote{This research was supported by the U.S. NSF under Grant No.
PHY-1414345.}}

\def\Title#1{\begin{center} {\Large #1 } \end{center}}
\def\Author#1{\begin{center}{ \sc #1} \end{center}}
\def\Address#1{\begin{center}{ \it #1} \end{center}}

\newcommand\pubblock{\rightline{\begin{tabular}{l} \pubnumber\\
         \pubdate  \end{tabular}}}
\newenvironment{Abstract}{\begin{quotation}  }{\end{quotation}}
\newenvironment{Presented}{\begin{quotation} \begin{center} 
             PRESENTED AT\end{center}\bigskip 
      \begin{center}\begin{large}}{\end{large}\end{center} \end{quotation}}
\def\Acknowledgements{\bigskip  \bigskip \begin{center} \begin{large}
             \bf ACKNOWLEDGEMENTS \end{large}\end{center}}

\input econfmacros.tex

\begin{document}
\begin{titlepage}
\pubblock

\vfill
\Title{Diagnosing New Physics with LUV and LFV B Decays.}
\vfill
\Author{ Alakabha Datta\support}
\Address{\napoli}
\vfill
\begin{Abstract}
 An important prediction of the  standard model is the universality of the gauge interactions of the three generation of  charged leptons. Violation of this universality would be a clean evidence of new physics (NP) beyond the standard model. In recent times
  anomalies in measurements  of certain $B$ decays indicate violation of lepton universality (LUV). I will discuss how one may  probe this LUV new physics via related decays and distributions. I will point out that
  LUV new physics can often lead to lepton flavor violation (LFV) and I will discuss  some promising decays to look for LFV new physics.
  \end{Abstract}
\vfill
\begin{Presented}
Thirteenth Conference on the Intersections of Particle and Nuclear Physics, CIPANP2018 \\
Palm Springs, CA, USA.
\end{Presented}
\vfill
\end{titlepage}
\def\thefootnote{\fnsymbol{footnote}}
\setcounter{footnote}{0}

\section{Introduction}
At present, there are several measurements of $B$ decays that may
indicate the presence of physics beyond the standard model (SM). In particular these measurements indicate  lepton universality violation (LUV). In the SM the gauge bosons couple equally to all members of the quark  and the  lepton families.  Hence  measurement of  LUV  is   evidence of physics beyond the SM.

The LUV   are observed in two groups- in charged current (CC) processes and in the neutral current (NC) processes. We will discuss them separately first and then consider how the CC and NC LUV might have a common origin.

\subsection{ { \bf { CC LUV}}}



The charged-current decays   $ B \to D^{(\ast)} \tau  \nu_\tau$   have been measured by the BaBar
  \cite{RD_BaBar}, Belle \cite{RD_Belle} and LHCb \cite{RD_LHCb}
  Collaborations. It is found that the values of the ratios
  $\RD \equiv {\cal B}(\bar{B} \to D^{(*)} \tau^{-}
  {\bar\nu}_\tau)/{\cal B}(\bar{B} \to D^{(*)} \ell^{-}
  {\bar\nu}_\ell)$ ($\ell = e,\mu$) considerably exceed their SM
  predictions. 
  These ratios of branching fractions
have certain advantages over the absolute branching fraction measurements
as they are  relatively less sensitive to form factor variations and several systematic uncertainties
such as those on the experimental efficiency as well as the dependence on the value of  $|V_{cb}|$ cancel in the ratios.
There are lattice determination of the ratio ${{R}}(D)_{SM}$    \cite{FNAL, HPQCD} that are in general agreement with one another.
\bea
{{R}}(D)_{SM} &=& 0.299 \pm 0.011, \quad \quad \mathrm{FNAL/MILC}\cite{FNAL} \nonumber \\
 {{R}}(D)_{SM} &=& 0.300\pm 0.008, \quad \quad \mathrm{HPQCD} \cite{HPQCD} \nonumber \\
{{R}}(D^\ast)_{SM}&=& 0.252\pm 0.003.
\eea
Calculation of ${R}(D^\ast)_{SM}$ is not available from lattice and so one has to use
 SM phenomenological prediction \cite{SM1, SM2, SM3} where the form factors are obtained from fits to the angular distributions in
 $\bar{B} \to D^{(*)} \ell^{-}
  {\bar\nu}_\ell$ .

By averaging the most recent measurements, the HFAG Collaboration has found \cite{HFAG}
\bea
{{R}}(D)  &=& 0.407 \pm 0.039 \pm 0.024, \nonumber \\
{{R}}(D^{\ast}) &=& 0.304 \pm 0.013 \pm 0.007,
\label{ratiotau}
\eea
where the first uncertainty is statistical and the second is
systematic. ${{R}}(D^{\ast})$ and ${{R}}(D)$ exceed the SM
predictions  by 3.4$\sigma$ and 2.3 $\sigma$, respectively.
The combined analysis of ${{R}}(D^{\ast})$ and ${{R}}(D)$, taking
into account measurement correlations, leads to a deviation
is 4.1$\sigma$ from the SM prediction \cite{HFAG}.


 In general there have been  
many analyses of  the   $\RD$ puzzles both in model independent framework as well as in specific models (see,  for example,
Refs.~\cite{RD1,RD2,RD3,RD4,RD5, RD6}). 


 \subsubsection {  Distributions and CP violation}
 The new physics  proposed for the $\RD$ puzzled can be probed in distributions \cite{ RD1, RD_dist1, RD_dist2}. Some of the  observables in the distributions have been measured. In the coming years more of  these will be measured.
 These measurements will discover or constrain new physics and will provide important clues to the nature of new physics. An important observable is CP violation in the distribution \cite{RD_dist1,Hagiwara} as this is free of hadronic uncertainties. Measurement of  non zero value of CP violation will be a clear sign of new physics. 
 \begin{figure}[htb!]
   \includegraphics[width=12.5cm]{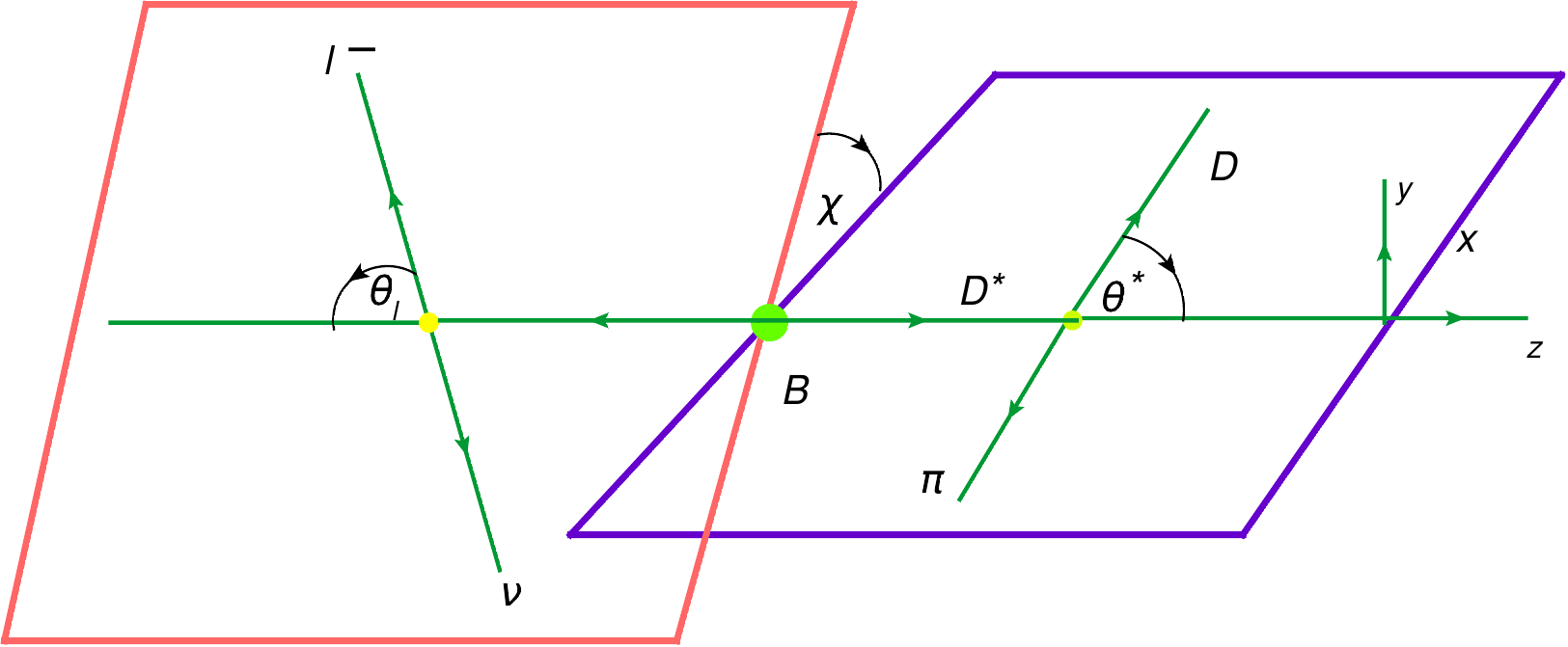}~\\
    \caption{The description of the angles $\theta_{\l{}}, \theta^*$ and $\chi$ in the angular distribution of $\bar{B}\rightarrow D^{*}  (\rightarrow D \pi)l^-\nu_l$  decay.}  
    \label{fig1} 
   \end{figure}
The complete three-angle distribution for the decay  $\bar{B}\rightarrow D^{*}  (\rightarrow D \pi)l^- \bar{\nu}_l$ in the presence of NP can be expressed in terms of four kinematic variables, $q^2$ ( the momentum transfer squared), two polar angles $\theta_l$, $\theta^{D^*}$, and the azimuthal angle $\chi$. The angle $ \theta_l$ is the polar angle between the charged lepton and the direction opposite to the $D^*$ meson in the $(l \nu_l)$ rest frame. The angle $\theta_{D^*}$  is the polar angle between the D meson and the direction of the $D^*$ meson in the  $(D \pi)$ rest frame. The angle $ \chi$  is the azimuthal angle between the two decay planes spanned by the 3-momenta of the $(D \pi)$ and $(l \nu_l)$ systems. These angles are described in Fig.~\ref{fig1}. The three-angle distribution can be obtained by using the helicity formalism.

\noindent We can write the angular distribution for   $\bar{B}\rightarrow D^{*}  (\rightarrow D \pi)l^- \bar{\nu}_l$    as  \cite{ RD1,  Dungel:2010uk, Richman:1995wm,Korner:1989qb,Korner:1989qa}
\bea
\label{3-foldAD_ex}
\frac{d^4\Gamma}{dq^2\, d\cos\theta_l\, d\cos\theta_{D^*}\, d\chi} & = &
 \frac{9}{32 \pi} NF   \Big(\sum^8_{i = 1} I_i + \frac{m_l^2}{q^2} \sum^8_{j = 1} J_i \Big), \nl
\eea
where the $I_i$ and $J_i$ are functions of the helicity amplitudes and the helicity angles \cite{RD_dist1}.

The complex NP couplings lead to CP violation which is sensitive to the angular terms
$\sin { \chi}$ and $\sin{2 \chi}$ in the distribution. The coefficients of these terms are 
triple products (TP)  and have the structure
$\sim Im[{\cal{A}}_i{\cal{A}}_j^*] \sim \sin( \phi_i-\phi_j) $, where 
${\cal{A}}_{i,j} =|{\cal{A}}_{i,j}|e^{ i \phi_{i,j}} $. In the SM these terms vanish, as there is only one dominant contribution to the decay and so all amplitudes have the same weak phase. 
Hence any non-zero measurements of TPs are clear signs of NP without any hadronic uncertainties.
For the charged conjugate modes, the weak phases change sign
and $\overline{{\cal{{A}}}_{i,j}} =|{\cal{A}}_{i,j}|e^{- i \phi_{i,j}} $ and  the TPs change sign. Even though we focus on $\tau$ final states, we should point out that this distribution is applicable also for $e$ and $\mu$ in the final state.
Since experiments have already studied the distributions for $e$, $\mu$ final states
it might be worth checking the $\sin { \chi}$ and $\sin{2 \chi}$ terms in the distributions for these decays for signals of non-SM physics.

 \subsubsection
{ {\bf {Other $b$ decays: $\lbt$}}
}
 There are other $b$ decays that can be used to  constrain the models discussed to explain the $\RD$ measurements.    It was pointed out that 
 the underlying quark level transition $\bctaunutau$ in the $\RD$ puzzles can be probed in both
$B$ and $\Lambda_b$ decays. 
The $\lbt$ decays could be useful to confirm
possible new physics in the $\RD$ puzzles and to point to the correct model of new physics.
Recently, in Ref.~\cite{Gutsche:2015mxa} this decay was discussed in the standard model and with new physics in Ref.~\cite{Wol, Shiv, Dutta}.
In Ref.~\cite{Shiv} the following quantities were calculated within the SM and with various new physics operators.  
\bea
\label{ratio1}
R_{\lamc} & = & \frac{BR[ \lbt]}{BR[\lbl]}, \
\eea
\bea
\label{ratio2}
B_{\lamc}(q^2) & = & \frac{\frac{d\Gamma[ \lbt]}{d q^2}}{\frac{ d \Gamma[\lbl]}{d q^2}}, \
\eea
where $\ell$ represents $\mu$ or $e$.
The value of  $R_{\lamc}$ in the SM  with lattice form factors can also be found in Ref.~\cite{Detmold:2015aaa}.
In a recent paper \cite{Saeed} the phenomenology of  this decay was studied in details with the most general new physics operators and  in specific models which have been considered for the $ \RD $ measurements. Lattice form factors for the $\Lambda_b \to \Lambda_c$ transitions were used  for precise calculations of the above ratios. In particular it was found that future measurements of the above ratios can strongly constrain or rule out certain models of new physics.
 
\subsubsection{ { \bf { NC LUV}}}
%

 The LHCb Collaboration has made measurements of $B
  \to K^* \mu^+\mu^-$ \cite{BK*mumuLHCb1,BK*mumuLHCb2} that deviate
  from the SM predictions \cite{BK*mumuSM}. The Belle Collaboration
  finds similar results \cite{BK*mumuBelle}. The main discrepancy is
  in the angular observable $P'_5$ \cite{P'5} though the significance of the
  discrepancy depends on the assumptions about the theoretical
  hadronic uncertainties.
The LHCb Collaboration has also measured the branching fraction and
performed an angular analysis of $B_s^0 \to \phi \mu^+ \mu^-$
\cite{BsphimumuLHCb1,BsphimumuLHCb2}. They find a $3.5\sigma$
disagreement with the predictions of the SM, which are based on
lattice QCD \cite{latticeQCD1,latticeQCD2} and QCD sum rules \cite{QCDsumrules}.

To find clear evidence of new physics one should consider observables largely free of hadronic uncertainties. One such observable is $R_K \equiv {\cal B}(B^+ \to K^+ \mu^+ \mu^-)/{\cal
  B}(B^+ \to K^+ e^+ e^-)$~\cite{RKtheory1, RKtheory2}, which has been measured by
 LHCb~\cite{RKexpt}:
\bea
R_K^\expt & = & 0.745^{+0.090}_{-0.074}~{\rm (stat)} \pm 0.036~{\rm (syst)} ~,~~ 1 \le q^2 \le 6.0 ~{\rm GeV}^2\,.
\label{RKexpt}
\eea
This differs from the SM prediction, $R_K^\SM = 1 \pm 0.01$~\cite{IsidoriRK} by $2.6\sigma$.
Note, the observable $R_K$ is a measure of lepton flavor universality and requires different new physics for the muons versus the electrons, while it is possible to explain the anomalies in the angular observables in $\bsmumu$ in terms of lepton flavor universal new physics~\cite{Datta:2013kja}.

Recently, the LHCb Collaboration reported the measurement of the
ratio $R_{K^*} \equiv {\cal B}(B^0 \to K^{*0} \mu^+ \mu^-)/{\cal
  B}(B^0 \to K^{*0} e^+ e^-)$ in two different ranges of the dilepton
invariant mass-squared $q^2$~\cite{RK*expt}:
\bea
R_{K^*}^\expt &  = &
\begin{array}{cc}
0.66^{+0.11}_{-0.07}~{\rm (stat)} \pm 0.03~{\rm (syst)} ~,~~ & 0.045 \le q^2 \le 1.1 ~{\rm GeV}^2 ~, \ \ \ (\rm{low} \ q^2)\\
0.69^{+0.11}_{-0.07}~{\rm (stat)} \pm 0.05~{\rm (syst)} ~,~~ & 1.1 \le q^2 \le 6.0 ~{\rm GeV}^2 ~, \ \ \ (\rm{central}\ q^2)\,.
\end{array}
\eea
These differ from the SM predictions
by 2.2-2.4$\sigma$ (low $q^2$) and 2.4-2.5$\sigma$
(central $q^2$), which  further strengthens the hint  of lepton
non-universality observed in $R_K$. 

Fits to new physics with heavy mediators have been considered by several groups but it has generally been difficult to understand the low $q^2$ measurements with heavy mediators \cite{datta_new_fit}.   There are attempts to understand the low $q^2$ $\RKstar$ measurement in terms of light mediators \cite{light_med, datta_LM2}. This scenario was found to have interesting signatures in coherent neutrino scattering\cite{coherent}.

 \subsubsection{ { \bf { Lepton Universality Violation and Lepton Flavor Violation}}}

%
%
Any interesting question is whether the  CC LUV and the NC LUV are related. In Ref.~\cite{AGC, RKRD}, it was pointed out that, assuming the scale of NP
is much larger than the weak scale,  operators contributing to $\RK$ anomalies   should be made invariant under the full
$SU(3)_C \times SU(2)_L \times U(1)_Y$ gauge group. There are two
possibilities if only left handed particles are involved:
\bea
{\cal O}_1^{NP} &=& \frac{G_1}{\Lambda_{NP}^2} ({\bar Q}'_L \gamma_\mu Q'_L) ({\bar L}'_L \gamma^\mu L'_L) ~, \nn\\
{\cal O}_2^{NP} &=& \frac{G_2}{\Lambda_{NP}^2} ({\bar Q}'_L \gamma_\mu \sigma^I Q'_L) ({\bar L}'_L \gamma^\mu \sigma^I L'_L) \nn\\
&=& \frac{G_2}{\Lambda_{NP}^2} \left[
2 ({\bar Q}'^{i}_L \gamma_\mu Q'^{j}_L) ({\bar L}'^{j}_L \gamma^\mu L'^{i}_L)
- ({\bar Q}'_L \gamma_\mu Q'_L) ({\bar L}'_L \gamma^\mu L'_L) \right] ~,
\label{NPoperators}
\eea
where $G_1$ and $G_2$ are both $O(1)$, and the $\sigma^I$ are the
Pauli matrices. Here $Q' \equiv (t',b')^T$ and $L' \equiv
(\nu'_\tau,\tau')^T$. The key point is that ${\cal O}_2^{NP}$ contains
both neutral-current (NC) and charged-current (CC) interactions. The
NC and CC pieces can be used to respectively explain the $R_K$ and
$R_{D^{(*)}}$ puzzles. In this scenario new physics affects only the third generation but via mixing effect LUV and lepton flavor violation(LFV) effects involving lighter generations are
generated \cite{GGL, RKRD}.


 In
Ref.~\cite{EffFT_3rdgen} 
 UV completions
that can give rise to ${\cal O}_{1,2}^{NP}$ [Eq.~(\ref{NPoperators})],
were discussed. These include leptoquark models and  vector boson (VB) models.
Concrete VB models are discussed as in Ref.~\cite{Isidori, Virto}.
Within specific models  there are new LUV processes as well as lepton flavor violating(LFV) processes.  A fairly exhaustive analysis of specific models was carried out in the recent publication\cite{datta_jhep_new}. Several decay modes were discussed which in the future could distinguish among the different models.
In particular it was shown that the observation of $\tau \to 3 \mu$ would be a clear sign of the VB model while the observation of $ \Upsilon \to \mu \tau$ would point towards the leptoquark model.


\section{Conclusion}
In conclusion we discussed the LUV anomalies in charged current (CC) and neutral current (NC) $B$ decays. We showed how related decays and angular distributions including CP violating observables can shed more light on the CC anomalies.
We discussed framework for a joint explanation of the CC and NC anomalies.  These scenarios also typically lead to lepton flavor violating decays and we identified
 the  interesting modes   $\tau \to 3\mu$ and  $\Upsilon(3S) \to \mu \tau$. Observations of these modes
will point to  specific models of new physics.

\Acknowledgements
I am grateful to  the high energy physics group at the University of California, Irvine for hospitality  where this work was completed.


\end{document}

%% file: econfmacros.tex
\def\beq{\begin{equation}}
\def\eeq#1{\label{#1}\end{equation}}
\def\eeqn{\end{equation}}

\def\beqa{\begin{eqnarray}}
\def\eeqa#1{\label{#1}\end{eqnarray}}
\def\eeqan{\end{eqnarray}}

\let\bar=\overbar

\def\Dslash{\not{\hbox{\kern-4pt $D$}}}
\def\dslash{\not{\hbox{\kern-2pt $\del$}}}

\def\msb{{\bar{\ssstyle M \kern -1pt S}}}

